\documentclass[12pt,a4paper]{article}
\usepackage{latexsym}
\usepackage{cite}
\DeclareSymbolFont{AMSb}{U}{msb}{m}{n}
\DeclareSymbolFontAlphabet{\Bbb}{AMSb}

\newcommand{\Z}{\Bbb{Z}}
\newcommand{\R}{\Bbb{R}}
\newcommand{\C}{\Bbb{C}}
\newcommand{\F}{\mathcal{F}}
\newtheorem{theorem}{Theorem}
\newtheorem{lemma}{Lemma}

\newcommand{\beq}{\begin{equation}}
\newcommand{\eeq}{\end{equation}}

\newcommand{\proof}{\textbf{Proof.\ }}
\newcommand{\qed}{\newline \hspace*{\fill} $\Box$ \par}

\begin{document}
\thispagestyle{empty}
\vspace*{-80pt} 
{\hspace*{\fill} Preprint-KUL-TF-99/10} 
\vspace{80pt} 
\begin{center} 
{\LARGE Mathematical Structure of Magnons in }\\[5pt]
 {\LARGE Quantum Ferromagnets  }
 \\[25pt]  
 
{\large  
    T.~Michoel\footnote{Aspirant van het Fonds voor Wetenschappelijk
    Onderzoek - Vlaanderen}\footnotetext{Email: {\tt tom.michoel@fys.kuleuven.ac.be}},
	A.~Verbeure\footnote{Email: {\tt andre.verbeure@fys.kuleuven.ac.be}}
    } \\[75pt]   
{Instituut voor Theoretische Fysica} \\  
{Katholieke Universiteit Leuven} \\  
{Celestijnenlaan 200D} \\  
{B-3001 Leuven, Belgium}\\[40pt]
\end{center} 
\begin{abstract}\noindent
We provide the mathematical structure and a simple, transparent and rigorous derivation
of the magnons as elementary quasi-particle excitations at low temperatures and in the
infinite spin limit for a large class of Heisenberg ferromagnets. The magnon canonical
variables are obtained as fluctuation operators in the infinite spin limit. Their quantum
character is governed by the size of the magnetization.
\end{abstract}
\newpage
\section{Introduction}
The appearance of spin waves in quantum ferromagnets at low temperatures is one of the
most basic physical quantum characteristics of quantum spin systems. It amounts to a boson
representation of the low temperature elementary excitations of a spin system. The basic
steps in the understanding of this phenomenon are done by Bloch \cite{01},
Holstein - Primakoff \cite{02}, van Kramendonk - van Vleck \cite{03}, and in the more
technical work by Dyson \cite{04, 05}.

From the point of view of mathematical physics, one discovers rigorously spin wave
properties at regular times as upper or lower bounds of correlations for low temperature
or ground states (see e.g. \cite{06}). Undoubtly, the so-called Bethe ansatz \cite{07} is
the most representative low temperature model of the spin wave theory, and, as is well
known, to prove or disprove that the Bethe ansatz is correct for some models is a serious
uptodate activity in mathematical physics. On the other hand, for a long time the spin
wave theory of Holstein - Primakoff itself calls for a simple, transparent and
mathematically rigorous setting, in the sense that the conditions for the appearance of
spin waves are clearly formulated, and that then the derivation of the spin waves or
magnons is rigorously obtained.

As far as we know, the most serious attempt for this has been made in \cite{08}. On the
basis of a classical domination principle which makes clear how to create the situation of
a unique ferromagnetic quantum ground state, these authors define a `physical'
Hamiltonian, which is of direct relevance to the analysis of the Holstein - Primakoff and
the Dyson formalism. They construct a `quadratic boson Hamiltonian' in configuration
space, and modulo an approximation, called plausible for large spins, they are able to
obtain satisfying results on the asymptotic equality of the free energies of the quadratic
boson Hamiltonian and the Heisenberg model in any dimension for low enough temperatures.

Here, stimulated by \cite{08}, we are able to clarify a number of aspects of the spin wave
or magnon theory which were still absent. As in \cite{08}, we limit ourself to
ferromagnets. Our main results are that we are able to use mathematical results on
non-commutative central limit theorems in order to scrutinize the large spin limit
correctly and to give a rigorous scheme for the formation of bosons. We are able to
perform this programme without any uncontrollable approximation. The result is that the
magnon canonical variables are nothing but fluctuation operators, whose mathematical
structure is developed in \cite{09, 10}. We perform rigorously the infinite spin limit, and 
prove that the Heisenberg model at low
temperature becomes a system of non-interacting magnons. We discuss the quantum character
of the magnons as a function of the magnetization at low temperatures.

Our results are on the level of the equilibrium states going beyond the study of its
thermodynamic properties. We create the
right conditions in order that the Heisenberg model system converges to a system of
magnons. All our conditions are in the direction of enough ferromagnetism. We are
convinced that results can be derived along the lines of what we have here also for
antiferromagnets if one is able to pin the suitable conditions on the interaction
constants in order to get specific antiferromagnetism.

\section{Low temperature magnons}
At each site $x$ of a finite domain $\Lambda$ of a cubic lattice $\Z^\nu$, consider the
 spin-$(2S+1)$ variables
$S^1(x), S^2(x), S^3(x)$, in the representation on $\otimes_{i=-S}^S \left(\C^2
\right)_i$ given by
\beq\label{def-S^mu}
S^\mu(x) = \sum_{i=-S}^S \sigma_i^\mu(x),
\eeq
$\mu=1,2,3$,  the $\sigma_i^\mu(x)$ Pauli matrices satisfying
\beq\label{Pauli-com-rel}
[\sigma_i^1(x),\sigma_j^2(y)]=2i\delta_{i,j}\delta_{x,y}\sigma_i^3(x), \; x,y\in\Z^\nu
\eeq
and its cyclic permutations of the components ($1,2,3$).
Let
\[
\sigma_i^\pm(x)=\frac{\sigma_i^1(x)\pm i\sigma_i^2(x)}{2}
\]
and
\[
S^\pm(x)=\sum_{i=-S}^S\sigma_i^\pm(x).
\]

The Heisenberg model Hamiltonian on $\Lambda$ is given by
\begin{eqnarray}\label{Heis-Ham}
H_\Lambda &=& -\frac{1}{2}\sum_{x,y\in\Lambda}\left\{ J(x,y)[S^1(x)S^1(y)+S^2(x)S^2(y)]
+ J_3(x,y) S^3(x)S^3(y) \right\} \nonumber\\
&&+ h\sum_{x\in\Lambda}S^3(x).
\end{eqnarray}
We assume finite range, translation invariant interactions, i.e. $J(x,y)=J(|x-y|)$, 
$J_3(x,y)=J_3(|x-y|)$ and $J(z)=J_3(z)=0$ for $z$ large enough, we assume also 
$J(x,x)=J_3(x,x)=0$. 

Remark that the representation of the spin variables is special in the sense that the
$S^\mu(x)$ (\ref{def-S^mu}), i.e. per lattice point $x$, are permutation invariant for
arbitrary permutations of the spin index $i=-S, -S+1, \ldots, S$. The Hamiltonian on the
other hand is not permutation invariant under permutations of the lattice indices
$x\in\Z^\nu$.

Conform to Bloch, Holstein - Primakoff, Dyson \cite{01,02,03,04,05}, we are interested in
the $S$ tending to infinity limit. In order to keep the model (\ref{Heis-Ham})
thermodynamically stable, one has to rescale it by the factor $2S+1$:
$H_\Lambda \to H_\Lambda^S=(2S+1)^{-1}H_\Lambda$. Applying a
rescaled magnetic field $h\to (2S+1)h$, we get
\begin{eqnarray*}
H_\Lambda^S &=& -\frac{1}{2(2S+1)}\sum_{x,y\in\Lambda}\left\{ J(x,y)[S^1(x)S^1(y)+S^2(x)S^2(y)]\right.
\\
&&\left.+ J_3(x,y) S^3(x)S^3(y) \right\} + h\sum_{x\in\Lambda}S^3(x).
\end{eqnarray*}
Or, rewritten, using $J(x,y)=J(y,x)$ and $J(x,x)=0$:
\begin{eqnarray}\label{H^S}
H_\Lambda^S &=& -\frac{1}{2(2S+1)}\sum_{x,y\in\Lambda}\left\{ 2J(x,y)[S^+(x)S^-(y)+S^-(x)S^+(y)]
\right.\nonumber\\
&&\left. + J_3(x,y) S^3(x)S^3(y) \right\} + h\sum_{x\in\Lambda}S^3(x)\nonumber\\
&=&-\frac{1}{2(2S+1)}\sum_{x,y\in\Lambda}\left\{ 4J(x,y)S^+(x)S^-(y)
+ J_3(x,y) S^3(x)S^3(y) \right\} \nonumber\\
&&+ h\sum_{x\in\Lambda}S^3(x).
\end{eqnarray}
We  are interested in the equilibrium states of this model in the $S\to\infty$ limit, and
in the thermodynamic limit $\Lambda\to\Z^\nu$. For each finite volume $\Lambda$, 
$H_\Lambda^S$ is permutation invariant for one-site observables $\sigma_i^\sharp(x)$ in 
spin space. Let $\omega$ be an equilibrium state, then per lattice site
$x\in\Lambda$, the state $\omega$ is again permutation invariant. In order to be able to
give a definite mathematical meaning to limits of the type, for one-site $x\in\Lambda$,
\[
\lim_{S\to\infty}\frac{S^\sharp(x)}{\sqrt{2S+1}},
\]
in the sense of non-commutative central limits, the state $\omega$ has to satisfy a
cluster property
\[
\lim_{i \to\infty}\omega\left( A_i(x)B_j(x)\right)=\omega(A)\omega(B),
\]
where $A(x)$ and $B(x)$ are products of the Pauli matrices (see e.g. \cite{09}). However
all clustering permutation invariant states are product states.
This implies that $\omega$, the infinite volume, infinite $S$ limit equilibrium state 
at inverse
temperature $\beta$ of $H_\Lambda^S$, is a product state in spin space:
\[
\omega(A_iB_j)=\omega(A)\omega(B)
\]
for $i\not=j$ and $A,B$ one-site observables in spin space, i.e. observables on 
$\otimes_{x\in\Lambda}(\C^2)_x$.

First we remark that for this product state in spin space, one has that in the
$\omega$-weak topology \cite{09}:
\[
weak-\lim_{S\to\infty}\left( \frac{1}{2S+1}\sum_{i=-S}^S \sigma_i^\sharp(x) \right) =
\omega(\sigma^\sharp(x))=\omega(\sigma^\sharp).
\]
The last equality follows from space translation invariance, which is a consequence of the
space translation invariance of (\ref{Heis-Ham}).

For $k\in\Lambda^* = \{ k= \frac{2\pi}{L}n\; ; \; n\in\Z^\nu \}$, $|\Lambda|=L^\nu$, let:
\[
\sigma_i^\pm(k)=\frac{1}{|\Lambda|^{1/2}}\sum_{x\in\Lambda}\left(\sigma_i^\pm(x)
-\omega(\sigma_i^\pm(x))\right)e^{\pm ik.x}
\]
\[
\sigma_i^3(k)=\frac{1}{|\Lambda|^{1/2}}\sum_{x\in\Lambda}\left(\sigma_i^3(x)
-\omega(\sigma_i^3(x))\right)e^{ik.x}
\]
and
\[
\tilde \sigma_i^\sharp(k)=\sigma_i^\sharp(k) + |\Lambda|^{1/2}\omega(\sigma^\sharp)\delta_{k,0}.
\]
Then
\beq\label{F_S}
F_{S}^\sharp(k)=\frac{1}{(2S+1)^{1/2}}\sum_{i=-S}^S \sigma_i^\sharp(k)
\eeq
are fluctuation operators in spin space and in volume space. The infinite $S$ limit
 of these operators is known to
exist due to the product state character of $\omega$ \cite{09}, i.e. the following
limits exist: for all $\lambda\in\R$,
\[
\lim_{S\to\infty}\omega\left( \exp\{i\lambda[F_S^\sharp(k)+F_S^\sharp(k)^*]\} \right)
\]
and
\[
\lim_{S\to\infty}\omega\left( \exp\{\lambda[F_S^\sharp(k)-F_S^\sharp(k)^*]\} \right).
\]
 The limit operators $\lim_{S\to\infty}F_S^\sharp(k)$ are denoted $F^\sharp(k)$,
they still depend on the volume $\Lambda$. It is straightforward to check that these
limits satisfy the canonical commutation relations.

Before taking the infinite $S$-limit, the operators
\beq\label{calF_S}
\F_S^\sharp(k)= \frac{1}{(2S+1)^{1/2}}\sum_{i=-S}^S \tilde \sigma_i^\sharp(k)
\eeq
can be used to rewrite $H_\Lambda^S$ (\ref{H^S}), after Fourier transformation, in the form
\begin{eqnarray*}
H_\Lambda^S&=&-\sum_{k\in\Lambda^*}\left\{ 2J(k)\F_S^+(k)\F_S^-(k) + \frac{1}{2}J_3(k)
\F_S^3(k)\F_S^3(-k)\right\}\\
&&+ h(|\Lambda|(2S+1))^{1/2}\F_S^3(0),
\end{eqnarray*}
where $J(k)=\sum_z J(z,0)e^{-ik.z}$, $J_3(k)=\sum_z J_3(z,0)e^{-ik.z}$.

The following Lemma characterizes the rotational invariance around the third axis, if
there is enough ferromagnetism.
\begin{lemma}\label{lemma1}
For $h$ sufficiently large, i.e. for
\[
h>\sum_{x\in\Lambda} [J_3(x,0)-J(x,0)] >0,
\]
any equilibrium state $\omega$ satisfies in the $\Lambda\to\Z^\nu$, $S\to\infty$ limit:
\[
\omega(\sigma^\pm)=0.
\]
\end{lemma}
\proof
Compute
\begin{eqnarray*}
&& [H_\Lambda^S, \frac{1}{|\Lambda|}\sum_{x\in\Lambda}\sigma_i^-(x)]\\
&=&\frac{-1}{(2S+1)|\Lambda|}\sum_{j=-S}^S \sum_{x,y}\left\{2J(x,y)\sigma_i^3(x)
\sigma_j^-(y) \right.\\
&&\left. - J_3(x,y)
[\sigma_i^-(x)\sigma_j^3(y)
+\sigma_j^3(y)\sigma_i^-(x)]\right\}
-\frac{2h}{|\Lambda|}\sum_x \sigma_i^-(x).
\end{eqnarray*}
Then the time invariance of $\omega$ implies
\begin{eqnarray*}
0 &=&\lim_{|\Lambda|\to\infty}\lim_{S\to\infty}
 \omega\left([H_\Lambda^S, \frac{1}{|\Lambda|}\sum_{x\in\Lambda}\sigma_i^-(x)]\right)\\
&=& -\lim_{|\Lambda|\to\infty}\frac{1}{|\Lambda|}\sum_{x,y}\left\{2J(x,y)\omega(\sigma^3)\omega(\sigma^-)
-2J_3(x,y)\omega(\sigma^3)\omega(\sigma^-)\right\}\\
&&-2h\omega(\sigma^-),
\end{eqnarray*}
or
\[
\omega(\sigma^-)\left\{\omega(\sigma^3)\sum_z[J_3(z,0)-J(z,0)]-h\right\}=0.
\]
Since $-1\leq\omega(\sigma^3)\leq 1$, taking $h>\sum_z[J_3(z,0)-J(z,0)]>0$ ensures
\[
\omega(\sigma^3)\sum_z[J_3(z,0)-J(z,0)]-h < 0.
\]
Hence
\[
\omega(\sigma^-)=0.
\]
\qed
This result means also that the operators $F^\pm_S$ and $\F^\pm_S$, as defined in
(\ref{F_S}) and (\ref{calF_S}) coincide:
\[
\F_S^\pm(k)=F_S^\pm(k),
\]
and hence also in the infinite $S$-limit:
\[
\F^\pm(k)=F^\pm(k).
\]

We compute the commutators for finite $S$:
\begin{eqnarray*}
[\F_S^+(k),\F_S^-(q)] &=&\frac{1}{|\Lambda|(2S+1)}\sum_{i=-S}^S\sum_x[\sigma_i^+(x),
\sigma_i^-(x)]e^{i(k-q).x}\\
&=&\frac{1}{|\Lambda|^{1/2}(2S+1)^{1/2}}\F_S^3(k-q),
\end{eqnarray*}
and:
\begin{eqnarray*}
[\F_S^3(k),\F_S^\pm(q)] &=&\frac{1}{|\Lambda|(2S+1)}\sum_{i=-S}^S\sum_x[\sigma_i^3(x),
\sigma_i^\pm(x)]e^{i(k\pm q).x}\\
&=&\pm\frac{2}{|\Lambda|^{1/2}(2S+1)^{1/2}}\F_S^\pm(q\pm k).
\end{eqnarray*}
Therefore the limits satisfy the boson commutation relations:
\beq\label{com-rel}
[\F^+(k),\F^-(q)]=\lim_{S\to\infty}[\F_S^+(k),\F_S^-(q)] = \omega(\sigma^3)\delta_{k,q},
\eeq
and
\beq\label{com-rel2}
[\F^3(k),\F^\pm(q)]=
\lim_{S\to\infty}[\F_S^3(k),\F_S^\pm(q)] = \pm 2\omega(\sigma^\pm)\delta_{k,q}=0,
\eeq
on the basis of Lemma \ref{lemma1}. 

The $\F^\pm(k)$ are the above rigorously defined magnon creation and annihilation
operators. Their existence and explicit properties are established as a straightforward
application of the non-commutative central limit theorems in \cite{09}, and the condition
of ferromagnetism in Lemma \ref{lemma1}.

Now we proceed by determining the equilibrium state $\omega$ completely. We use the
definition of equilibrium state by means of correlation inequalities \cite{11}, i.e. for
temperatures $T>0$ ($\beta=\frac{1}{kT}<\infty$), a state $\omega$ is an equilibrium state
if and only if it satisfies the energy-entropy balance inequalities:
\beq\label{cor-eq}
\lim_{\Lambda\to\infty}\lim_{S\to\infty}\beta\omega\left( X^*[H_\Lambda^S,X] \right)
\geq \omega(X^*X)\ln\frac{\omega(X^*X)}{\omega(XX^*)},
\eeq
for all local observables $X$. We prove:
\begin{theorem}
In the ferromagnetic region (see Lemma \ref{lemma1}), in the infinite $S$-limit, the
equilibrium state $\omega$ of the Heisenberg model is a quasi-free state on the
fluctuation operators algebra, generated by the $\{ \F^\sharp(k), k\in [0,2\pi]^\nu \}$,
completely determined by the two-point function
\beq\label{2-pnt-f}
\omega\left( \F^+(q)\F^-(q) \right)=\frac{-\omega(\sigma^3)}{e^{2\beta[-\omega(\sigma^3)
(J_3(0)-J(q))+h]}-1}.
\eeq
\end{theorem}
\proof
Take first $X=\F_S^-(q)$, and compute
\begin{eqnarray*}
&&[H_\Lambda^S,\F_S^-(q)]\\
&=&-\sum_k  2J(k)[\F_S^+(k),\F_S^-(q)]\F_S^-(k)\\
&& -\sum_k\frac{1}{2}J_3(k)\left( [\F_S^3(k),\F_S^-(q)]\F_S^3(-k)
+ \F_S^3(k)[\F_S^3(-k),\F_S^-(q)]\right) \\
&&+ h |\Lambda|^{1/2}(2S+1)^{1/2} [\F_S^3(0),\F_S^-(q)]\\
&=& -\frac{2}{|\Lambda|^{1/2}(2S+1)^{1/2}}\sum_k J(k)\F_S^3(k-q)\F_S^-(k)\\
&& + \frac{1}{|\Lambda|^{1/2}(2S+1)^{1/2}}\sum_k J_3(k) \left( \F_S^-(q-k)\F_S^3(-k)+
\F_S^3(k)\F_S^-(q+k)\right)\\
&&-2h\F_S^-(q).
\end{eqnarray*}

Then:
\begin{eqnarray*}
&&\beta\omega\left( \F_S^+(q)[H_\Lambda^S,\F_S^-(q)] \right)\\
&=&-\frac{2\beta}{|\Lambda|^{1/2}(2S+1)^{1/2}}\sum_k J(k)\omega\left(
\F_S^+(q)\F_S^3(k-q)\F_S^-(k) \right)\\
&&+ \frac{\beta}{|\Lambda|^{1/2}(2S+1)^{1/2}}\sum_k J_3(k) \omega\left(
\F_S^+(q) \F_S^-(q-k)\F_S^3(-k)\right)\\
&&+\frac{\beta}{|\Lambda|^{1/2}(2S+1)^{1/2}}\sum_k J_3(k)\omega\left(
\F_S^+(q)\F_S^3(k)\F_S^-(q+k)\right)\\
&&-2\beta h \omega\left( \F_S^+(q)\F_S^-(q)\right).
\end{eqnarray*}

Use that for $A,B$ one-site observables in spin space such that $\omega(A)=0$ (see
\cite{09}):
\[
\lim_{S\to\infty}\frac{1}{(2S+1)^{1/2}}\omega(\F_S(A)^*\F_S(B)\F_S(A))=\omega(A^*A)\omega(B),
\]
to calculate
\begin{eqnarray*}
&&\lim_{S\to\infty}\beta\omega\left( \F_S^+(q)[H_\Lambda^S,\F_S^-(q)] \right)\\
&=&-\frac{2\beta}{|\Lambda|^{1/2}}\sum_k J(k)\lim_{S\to\infty}\omega\left(
\F_S^+(q)\F_S^-(k) \right) \omega(\sigma^3)\delta_{k,q}|\Lambda|^{1/2}\\
&&+\frac{\beta}{|\Lambda|^{1/2}}\sum_k J_3(k)\lim_{S\to\infty}\omega\left(
\F_S^+(q)\F_S^-(q-k) \right) \omega(\sigma^3)\delta_{k,0}|\Lambda|^{1/2}\\
&&+\frac{\beta}{|\Lambda|^{1/2}}\sum_k J_3(k)\lim_{S\to\infty}\omega\left(
\F_S^+(q)\F_S^-(q+k) \right) \omega(\sigma^3)\delta_{k,0}|\Lambda|^{1/2}\\
&&-2\beta h \lim_{S\to\infty}\omega\left( \F_S^+(q)\F_S^-(q)\right),
\end{eqnarray*}
and
\[
\lim_{S\to\infty}\omega\left( \F_S^+(q)\F_S^-(q)\right)=
\omega\left( \F^+(q)\F^-(q)\right).
\]
Then
\[
\lim_{\Lambda\to\infty}\lim_{S\to\infty}\beta\omega\left( \F_S^+(q)[H_\Lambda^S,\F_S^-(q)] \right)
\]
\[
=2\beta\left\{\omega(\sigma^3)[J_3(0)-J(q)]- h\right\}\omega\left( \F^+(q)\F^-(q)\right).
\]

After substitution into the correlation inequality (\ref{cor-eq}), one gets:
\[
-2\beta\left\{-\omega(\sigma^3)[J_3(0)-J(q)]+ h\right\} \geq 
\ln\frac{\omega\left( \F^+(q)\F^-(q)\right)}{\omega\left( \F^-(q)\F^+(q)\right)}.
\]

Interchanging the role of $\F^+(q)$ and $\F^-(q)$, i.e. take now $X=\F^+_S(q)$ in
(\ref{cor-eq}) and repeat the computation above, then:
\[
2\beta\left\{-\omega(\sigma^3)[J_3(0)-J(q)]+ h\right\} \geq 
\ln\frac{\omega\left( \F^-(q)\F^+(q)\right)}{\omega\left( \F^+(q)\F^-(q)\right)}
\]

These two inequalities, combined with the commutation relation (\ref{com-rel}) yield
\[
\ln\frac{\omega\left( \F^+(q)\F^-(q)\right)-\omega(\sigma^3)}
{\omega\left( \F^+(q)\F^-(q)\right)} =
2\beta\left\{-\omega(\sigma^3)[J_3(0)-J(q)]+ h\right\},
\]
or the expected two-point function
\[
\omega\left( \F^+(q)\F^-(q)\right)=\frac{-\omega(\sigma^3)}
{e^{2\beta\left\{-\omega(\sigma^3)[J_3(0)-J(q)]+ h\right\}}-1}.
\]

Finally if one takes for $X$ higher order monomials in the $\F_S^\sharp(q)$, one derives
readily also from (\ref{cor-eq}), that the higher order point correlation functions are
sums of products of this two-point function, proving that the state $\omega$ is
quasi-free. As this amounts to a straightforward computation, we leave it as an exercise
for the reader.
\qed

The basic two-point function (\ref{2-pnt-f}) contains still the magnetization
$\omega(\sigma^3)$. Using (\ref{Pauli-com-rel}) and (\ref{calF_S}), one gets a
self-consistency equation for the magnetization
\beq\label{selfcons}
\frac{1}{|\Lambda|}\sum_{q\in\Lambda^*}\omega\left( \F^+(q)\F^-(q)\right)
=\frac{1+\omega(\sigma^3)}{2}.
\eeq
Let
\beq\label{D} 
D(x,y)=\lambda(x)\delta_{x,y}-J(x,y), 
\eeq
where
\[
\lambda(x)=\sum_y J_3(x,y)
\]
and suppose that we limit ourself to the {\em ferromagnetic situation} (e.g.
$J_3(x,y)\geq |J(x,y)|$, see \cite{08}), expressed by the condition that the matrix $D$
(\ref{D}) is positive definite:
\[
D(x,y)\geq 0.
\]
This implies
\[
D(q)=\sum_z D(z,0) e^{-iq.z}=J_3(0)-J(q) \leq \sum_z D(z,0)=D(0).
\]
Hence
\[
h > \sum_z [J_3(z,0)-J(z,0)]=D(0)\geq D(q)
\]
and
\[
h-\omega(\sigma^3)[J_3(0)-J(q)]=h-\omega(\sigma^3)D(q)\geq h-D(q) \geq h-D(0)>0.
\]

From this, and from (\ref{2-pnt-f}), it follows that
\[
0\leq\omega\left( \F^+(q)\F^-(q)\right)=\frac{-\omega(\sigma^3)}
{e^{2\beta\left\{h-\omega(\sigma^3)D(q)\right\}}-1}\leq \frac{1}{e^{2\beta h}-1}.
\]
The first inequality yields $-1\leq\omega(\sigma^3)\leq 0$.

Using (\ref{selfcons}), one gets
\begin{eqnarray}
\omega(\sigma^3)&\leq& -1 + \frac{2}{e^{2\beta h}-1}\\
&\leq& -1 + \frac{2}{e^{2\beta D(0)}-1} 
\simeq -1 + 2 e^{-2\beta D(0)},
\end{eqnarray}
establishing a bound on the magnetization for low temperatures as a function of the
interaction constants. The bound measures the deviation of the magnetization from its
ground state value, equal to $-1$, for small temperatures.

Remark that for $\omega(\sigma^3)=-1$, the ground state ($\beta\to\infty$) value of the
ferromagnetic system, the magnon creation and annihilation operators form a bosonic pair
satisfying the canonical commutation relations
\[
[\F^-(q),\F^+(q)]=1,
\]
and that the ground state $\omega$ is a magnon Fock state
\[
\omega\left(\F^+(q)\F^-(q)\right)=\lim_{\beta\to\infty}\frac{1}{e^{2\beta[D(q)+h]}-1}=0.
\]

In general, for all temperatures, the magnetization $\omega(\sigma^3)$ plays the role of
the quantization parameter (a Planck's constant) (see (\ref{com-rel})) for the field of
magnons. All quantum character of the magnons vanishes if one chooses the magnetic field
$h$ and/or the interaction constants ($D(q)$) and/or the temperature such that the
magnetization vanishes.

Concerning the magnetization fluctuation operators $F_S^3(q)$, the ferromagnetic
conditions ($h-D(0)>0$, $D\geq 0$) are such that its infinite $S$-limit $F^3(q)$ commutes
with all other magnon observables (see (\ref{com-rel2})). They become classical
observables, and disappear completely from the action of the system Hamiltonian. It does
not mean that there are no magnetization fluctuations.

The original system with Hamiltonian $H_\Lambda^S$ in terms of the fluctuation
observables $\{\F_S^\pm(q),F^3_S(q)\}_{q\in\Lambda^*}$, becomes in the infinite $S$-limit
a system of non-interacting magnons with Hamiltonian
\beq\label{q-magnHam}
\mathcal{H}_\Lambda=\sum_{q\in\Lambda^*}\epsilon(q)\F^+(q)\F^-(q),
\eeq
where the magnon fluctuation creation and annihilation operators satisfy
\[
[\F^-(q),\F^+(q')]=-\omega(\sigma^3)\delta_{q,q'},
\]
and where the spectrum is given by
\[
\epsilon(q)=2\left(J_3(0)-J(q)+ \frac{h}{-\omega(\sigma^3)}\right).
\]
Remark that if $min(h,D(0))>0$, then already $\epsilon(q=0)>0$, i.e. there is no condensation of
magnons in the zero mode $q=0$. In particular, this is the case under our assumptions.

By inverse Fourier transform, i.e.
\[
\F^\pm(x)=\frac{1}{|\Lambda|^{1/2}}\sum_{q\in\Lambda^*}\F^\pm(q)e^{\mp iq.x},
\]
(\ref{q-magnHam}) can be written in configuration space:
\beq\label{x-magnHam}
\mathcal{H}_\Lambda=2\sum_{x,y}D(x,y)\F^+(x)\F^-(y)+\frac{2h}{-\omega(\sigma^3)}
\sum_x \F^+(x)\F^-(x),
\eeq
with
\beq\label{x-comrel}
[\F^-(x),\F^+(y)]=-\omega(\sigma^3)\delta_{x,y}
\eeq
and $D$ is as defined above. Written this way, it is clear that $\mathcal{H}_\Lambda$ is 
nothing but the `quadratic boson Hamiltonian' of \cite{08}. 

Finally we look for the dynamics of the magnon excitation number operator
$\F^+(x)\F^-(x)$. On the basis of (\ref{selfcons}) the expectation value of this operator
is related to the magnetization. Therefore the time evolution of this number operator is
related to the time evolution of the magnetic moment in a ferromagnet.

The equation of motion for $\F^+(x)\F^-(x)$ is given by:
\begin{eqnarray}\label{dyneq}
&& \frac{\partial}{\partial t} \F^+(x)\F^-(x) \nonumber\\ &=& i\lbrack
\mathcal{H}_\Lambda, \F^+(x)\F^-(x)\rbrack \nonumber\\
&=&-2\omega(\sigma^3)\sum_y J(x,y) \left(\F^+(y)\F^-(x) 
- \F^-(y)\F^+(x)\right).
\end{eqnarray}
Indeed, using (\ref{x-comrel}) and $J(x,x)=0$ yields straightforwardly the result.

This dynamical equation (\ref{dyneq}) can be compared with the macroscopic equation of
motion for the magnetic moment, first derived in \cite{12} (see also \cite{13,14}). A
recent rigorous derivation of this equation for zero temperature is given in \cite{15},
applying a hydrodynamical limit or a Lebowitz - Penrose approximation \cite{16}. Remark
that equation (\ref{dyneq}) however, is valid for all temperature equilibrium states.


\begin{thebibliography}{abcd}
\bibitem{01} F. Bloch; Z. Physik {\bf 61}, 206 (1930).

\bibitem{02} T. Holstein, H. Primakoff; Phys. Rev. {\bf 58}, 1098 (1940).

\bibitem{03} J. van Kranendonk, J.H. van Vleck; Rev. Mod. Phys. {\bf 30}, 1 (1958).

\bibitem{04} F.J. Dyson; Phys. Rev. {\bf 102}, 1217 (1956).

\bibitem{05} F.J. Dyson; Phys. Rev. {\bf 102}, 1230 (1956).

\bibitem{06} F.J. Dyson, E.H. Lieb, B. Simon; J. Stat. Phys. {\bf 18}, 335 (1978).

\bibitem{07} H.A. Bethe; Z. Physik {\bf 71}, 205 (1931).

\bibitem{08} J.L. van Hemmen, A.A.S. Brito, W.F. Wreszinski; J. Stat. Phys. {\bf 37}, 187
(1984)

\bibitem{09} D. Goderis, A. Verbeure, P. Vets; Prob. Th. Rel. Fields {\bf 82}, 527 (1989).

\bibitem{10} D. Goderis, A. Verbeure, P. Vets; Commun. Math. Phys. {\bf 128}, 533 (1990).

\bibitem{11} M. Fannes, A. Verbeure; Commun. Math. Phys. {\bf 55}, 125 (1977); {\bf 57},
165 (1977).

\bibitem{12} L.D. Landau, E.M. Lifshitz; Phys. Zeitschrift der Sowjetunion {\bf 108}, 153,
(1935); Statistical Physics, Vol. 9.

\bibitem{13} A.I. Akhiezev, V.G. Bar'yakhtar, S.V. Peletminskii; Spin waves, in Low
Temperature Physics, Vol. 1, North-Holland Amsterdam (1968).

\bibitem{14} C. Herring, C. Kittel; Phys. Rev. {\bf 81}, 869 (1951).

\bibitem{15} M. Moser, A. Prets, W.L. Spitzer; Time evolution of spin waves in
ferromagnets at zero temperature; preprint ThPh-Vienna, Sept. 1998.

\bibitem{16} J.L. Lebowitz, O. Penrose; J. Math. Phys {\bf 7}, 98 (1966).
\end{thebibliography}
\end{document}